\documentclass[]{spie}  

\usepackage{amsmath,amsfonts,amssymb}
\usepackage{graphicx}
\usepackage{float}
\usepackage[rightcaption]{sidecap}
\usepackage{lineno}
\usepackage{xcolor}
\usepackage{soul}
\usepackage{multirow}  
\usepackage{geometry}  
\begin{document} 

\title{PRIMAger General Observer programs: a {\Large $\pi$}-sr Infrared Survey and other General Observer wide-field programs}

\author[a]{Denis Burgarella}
\author[b]{Matthieu B\'ethermin}
\author[a]{Alessandro Boselli}
\author[c]{J. M. S. Donnellan}
\author[d]{C. Darren Dowell}
\author[a]{G. Lagache}
\author[c]{Seb Oliver}
\author[e]{Herv\'e Dole}

\affil[a]{Aix Marseille Univ, CNRS, CNES, LAM, Marseille, France}
\affil[b]{Universit\'e de Strasbourg, CNRS, Observatoire astronomique de Strasbourg, UMR 7550, 67000 Strasbourg, France}
\affil[c]{Astronomy Centre, University of Sussex, Falmer, Brighton BN1 9QH, UK}
\affil[d]{Jet Propulsion Laboratory, California Institute of Technology, 4800 Oak Grove Drive, Pasadena, CA 91109, USA}
\affil[e]{Universit\'e Paris-Saclay, CNRS, Institut d'astrophysique spatiale, 91405, Orsay, France}

\authorinfo{Further information: (Send correspondence to Denis Burgarella)\\E-mail: denis.burgarella@lam.fr}

\pagestyle{plain} 
\setcounter{page}{1} 
 
\maketitle

\begin{abstract}
The PRobe far-Infrared Mission for Astrophysics (PRIMA) is a cryogenically-cooled, far-infrared (far-IR) observatory expected to begin serving the astronomical community by early 2030. The mission features two advanced instruments: PRIMAger and FIRESS. PRIMAger will operate across the mid- to far-IR spectrum, covering wavelengths from approximately 25 to 260 $\mu$m. It will offer hyperspectral imaging in medium resolution bands (R $\sim$ 8, using a linear variable filter) from 25 to 80 $\mu$m, and broad band (R $\sim$ 4) photometric and polarimetric imaging in four bands spanning 80 to 260 $\mu$m.

The capabilities of PRIMAger will enable a broad range of unique scientific programs, accessible through General Observer (GO) projects. 

In this paper, we present and define a PRIMAger survey over 25\% of the sky, called $\pi$-IR survey. This survey would exploit PRIMAger's hyperspectral and polarimetric modes to collect data on about 8 $\times$ 10$^{6}$ galaxies to z $\sim$ 4. The R=8 spectral resolution of the PRIMAger Hyperspectral Imaging (PHI) filters will enable users to study the emission of polycyclic aromatic hydrocarbon (PAH). A large sample of galaxies will be observed with the polarimetric bands of PRIMAger, allowing unique statistical information for galaxies to be harvested for the first time. 

\end{abstract}


\section{A short introduction to PRIMA and PRIMAger}
\label{sec:whatwho}

PRIMA (Glenn et al., JATIS ref) features a cryogenically 4.5-K cooled 1.8-m diameter IR telescope. PRIMA is designed to carry two science instruments that enable ultra-high sensitivity imaging (PRIMAger, Ciesla et al. JATIS ref) and spectroscopic studies (FIRESS, Bradford et al. JATIS refs) in the 25 to 264 $\mu$m wavelength range. The resulting observatory is a powerful survey and discovery machine.

\begin{figure} [H]
   \includegraphics[width=16.5cm]{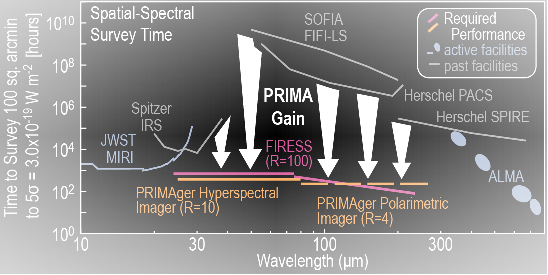}
  \caption{PRIMA will offer a huge gain in sensitivity with respect to previous mid- and far-IR facilities. PRIMA also perfectly fills the 25 to 260 $\mu$m wavelength range between JWST and ALMA/NOEMA. In addition to the hyperspectral imaging capabilities (PRIMAger Hyperspectral Imaging: PHI) with the Linear Variable Filter in confusion-limited bands from 25 to 80 $\mu$m, PRIMAger's polarimetric capabilities from about 80 to 260 $\mu$m (PRIMAger Polarimetric Imaging: PPI) will provide a powerful instrument to the entire astrophysical community at the beginning of the next decade.}
  \label{fig:mapingspeed}
\end{figure}

Fig.~\ref{fig:mapingspeed} shows the expected gain of 3 - 5 orders of magnitude in the spectral mapping speed. This huge gain provides us with an exceptional science discovery space in between JWST at shorter wavelengths and ALMA/NOEMA at longer wavelengths. We stress that PRIMA will be a community facility: about 75\% of the total observing time will be allocated to the General Observer (GO) program. The community interest in using PRIMA for GO programs is very strong. The PRIMA GO Book Volume-1 [\citenum{Moullet2023}], collects, lists, and describes 76 possible science programs. A second PRIMA GO Book Volume-2 will present about 120 GO cases. The proposed PRIMAger contributions cover all astrophysical sub-domains: compact objects and energetic phenomena, cosmology, galaxies, exoplanets, exobiology and Solar System, interstellar medium (ISM), stellar and planetary system formation, stars. Several very wide-field or even all-sky surveys focusing on various astrophysical topics will be included in this PRIMA GO Book Volume-2.

Within the PRIMA GO Book Volume-1, about one-third of proposed programs make use of PRIMAger only, and another third combine PRIMAger and FIRESS. PRIMAger is thus a fundamental part of the PRIMA project that will explore the cosmos in the mid- to far-IR wavelength range from space, and harvest unique data that cannot be collected by any other facility presently on the sky, yielding enormous benefits to the entire astrophysical community. 

\section{The need for an IR versatile space imager}
\label{sec:NeedforIR}

Astrophysics is fundamentally a multi-wavelength science. The wide variety of physical processes in astronomical objects can only be fully understood by probing the electromagnetic spectrum from $\gamma$-rays to radio. Our ability to make scientific progress thus depends critically on maintaining access to the full spectral energy distribution (SED). Although ground-based observatories effectively cover the optical, near-IR, and radio windows, the mid- to far-infrared (25–260\,\textmu m) remains accessible only from space.

The astronomical IR legacy has been defined by a series of landmark missions:
\begin{itemize}
    \item \textbf{IRAS} [\citenum{Neugebauer1984}] was the first to carry out an all-sky IR survey at 12, 25, 60, and 100 \textmu m.
    \item \textbf{ISO} [\citenum{Kessler2003}], operated by ESA from 1995 to 1998, studied the 2.5–240 \textmu m range with a 60\,cm cooled telescope.
    \item \textbf{AKARI} [\citenum{Murakami2007}], launched in 2006 by JAXA with international collaboration, operated from 1.7 to 180 \textmu m with a 68.5 cm telescope cooled to 6 K.
    \item \textbf{Spitzer} [\citenum{Werner2004}], active from 2003 to 2020, observed from 3.6 to 160 \textmu m with an 85 cm mirror cooled to 5.5 K.
    \item \textbf{WISE} [\citenum{Wright2010}], launched in 2009 by NASA, surveyed the sky at 3.4, 4.6, 12, and 22 \textmu m with a 40 cm telescope.
    \item \textbf{Herschel} [\citenum{Pilbratt2010}], ESA’s far-IR flagship (2009–2013), featured a 3.5 m mirror at 80~K and covered the 55–672 \textmu m range.
    \item \textbf{SOFIA} [\citenum{Young2012}], the Stratospheric Observatory For Infrared Astronomy (SOFIA) is an airborne observatory consisting of a specially modified Boeing with a 2.7 m telescope, flying at 13.7 km and designed to observe at wavelengths from 0.3 \textmu m to 1.6 mm, and with polarimetry being a key capability. 
    \item \textbf{JWST} [\citenum{Rigby2023}], The James Webb Space Telescope (JWST) is a 6.6m, $<$50 K, IR-optimized (from about 0.6 to 28.3 \textmu m) observatory, that is now in science operations. The project is an international collaboration among NASA, the European Space Agency (ESA), and the Canadian Space Agency (CSA).
\end{itemize}

Despite this rich history, ESA’s decision to cancel the cryogenically-cooled SPICA mission leaves astronomy without any active or planned observatory for the mid- to far-IR for at least the next one or two decades.

PRIMAger onboard the 4.5 K-cooled PRIMA telescope is designed to restore and extend our capabilities in this domain. It offers (also see Tab.~\ref{tab:prima_instruments}):
\begin{itemize}
    \item A short-wavelength (26–80 \textmu m) hyperspectral imager ($R \approx 8$),
    \item A long-wavelength (80–260 \textmu m) multi-band polarimetric imager ($R \approx 4$),
    \item Simultaneous full-range observation across all bands,
    \item $\sim$4 arcsec angular resolution at the shortest wavelengths,
    \item Extremely fast mapping speed (see Fig.~\ref{fig:mapingspeed}).
\end{itemize}

\begin{table}[htbp]
\centering
\resizebox{\textwidth}{!}{%
\begin{tabular}{l|cc|ccccc}
\hline
\textbf{Parameter} & \multicolumn{2}{c|}{\textbf{PRIMA Hyperspectral Imager}} & \multicolumn{4}{c|}{\textbf{PRIMA Polarimetry Imager}} \\
\hline
 & \textbf{PHI1} & \textbf{PHI2} & \textbf{PPI1} & \textbf{PPI2} & \textbf{PPI3} & \textbf{PPI4} \\
\hline
\textbf{Wavelength ($\mu$m)} & 24--45 & 45--84 & 92 & 126 & 183 & 235 \\ 
\textbf{Spectral resolving power} & 8 & 8 & 4 & 4 & 4 & 4 \\ 
\textbf{Polarimetry} & - & - & Yes & Yes & Yes & Yes \\ 
\textbf{Band centre FWHM ($''$)} & 4.7 & 8.7 & 10.9 & 14.9 & 21.7 & 27.6 \\ 
\textbf{Pixel count} & $61\times24$ & $34\times14$ & $34\times29$ & $25\times21$ & $18\times15$ & $14\times12$ \\ 
\textbf{Field of view} & $3.9'\times1.3'$ & $3.9'\times1.4'$ & $4.2'\times4.2'$ & $4.2'\times4.2'$ & $4.2'\times4.2'$ & $4.2'\times4.1'$ \\ \hline
\end{tabular}
}
\caption{Key instrument parameters for the PRIMA Hyperspectral and Polarimetry Imagers. The table lists wavelength coverage, resolving power, polarimetric capability, spatial resolution, array format, pixel size, and field of view for each module. Data sourced from the official PRIMA instrument page at https://prima.ipac.caltech.edu/page/instruments.}
\label{tab:prima_instruments}
\end{table}

This configuration makes PRIMAger the first mission to combine broad spectral coverage, high spatial resolution, polarimetry, and wide-area survey capability, a major advance in infrared space instrumentation.

\subsection*{Wide-Field Survey Power: Rare and Distant Objects}

PRIMAger is uniquely optimized for large-scale surveys. Its combination of fast mapping speed, large instantaneous field of view, and simultaneous multi-band capability enables ambitious programs to systematically map the sky in IR light. In particular, it could support a “wedding-cake” strategy with wide-and-shallow surveys for rare, bright objects (e.g., luminous infrared galaxies, strongly lensed systems), and deep-and-narrow surveys to uncover faint, high-redshift sources such as dust-obscured galaxies in the epoch of reionization. This flexibility enables PRIMAger to deliver both the breadth to characterize populations statistically and the depth to reveal the underlying physical mechanisms.

No previous mission has combined this level of spatial resolution, spectral coverage, and mapping efficiency. PRIMAger thus stands out as a true survey instrument, capable of building reference IR sky maps for decades to come.

\subsection*{Infrared Polarimetry: Mapping the Magnetic Universe}

Equally transformative is PRIMAger’s capability for far-IR polarimetry. Dust grains aligned with magnetic fields emit polarized thermal radiation, which carries vital information about the structure and role of magnetic fields in galaxies and the ISM. PRIMAger will be the first mission to deliver:
multi-band polarimetric maps in the 80–260 \textmu m range.

Although Planck provided all-sky polarization maps at much lower resolution, and ALMA offers exquisite detail but in tiny fields, PRIMAger will bridge these scales, enabling multi-scale studies of magnetism from the diffuse ISM to dense star-forming cores, across a variety of environments.

\subsection*{A Legacy Dataset for the Community}

As IRAS, Spitzer, and Herschel built enduring scientific legacies, PRIMAger is poised to generate the most comprehensive mid- to far-IR survey ever conducted. With simultaneous imaging and polarimetry across a broad spectral range and wide fields, PRIMAger will deliver a legacy dataset that will serve as a foundation for infrared astrophysics for the coming decades.

In this paper, we present a non-exhaustive overview of science potential of PRIMAger. Among the most ambitious prospects is a wide-area survey covering about 25\% of the sky (see Sect.~\ref{sec:allsky}), delivering a legacy dataset that offers not just fluxes, but also spectral and polarimetric insights into the dusty, magnetized universe.

\section{A {\Large $\pi$-IR} survey}
\label{sec:allsky}

The science of PRIMAger is supported by a strong tripod where the IR coverage and polarimetric capabilities are complemented with high mapping speed. This high mapping speed is a strong asset for PRIMA that could even enable an all-sky survey with PRIMAger with the hyperspectral and polarimetric modes. This should be viewed in the context of the number of proposed large programs which require several tens and even hundreds of degrees$^2$, with hundreds of hours of exposure time. 

In a previous project, Wright et al. [\citenum{Moullet2023}] describe a PPI (polarimetry: 80 - 260 $\mu$m) all-sky survey. The present {$\pi$-IR} survey will restrict the observations to about 25 \% of the sky: a $\pi$-IR survey. When observing with PPI, PHI (hyperspectral: 20 - 80 $\mu$m) observations are taken simultaneously. Such a survey will detect many new sources for GO programs to target for PRIMA follow-up observations. A very wide survey that covers a substantial part of the sky is the best way to circumvent any bias due to limited statistics. Moreover, this survey will revisit the Ecliptic Poles at every rotation, providing a "free" deep field, overlapping with the JWST time domain field, the Hubble deep fields, and other deep fields by Planck, eROSITA, and WISE, and enhancing their science. For comparison, in the WISE survey [\citenum{Schlafly2019}] at low ecliptic  latitude, the typical number of observations in the unWISE coadds is about 120 per band. On the other hand, at the ecliptic poles, this number reaches 23,000. Even though the size of the instantaneous field of view and the size of individual tiles are important factors, this suggests that the cumulative exposure time would lead to an improvement of a factor of $\sim$16 in limiting flux in the deeper fields at the ecliptic poles compared to a single pointing.

For such a survey, Wright et al. [\citenum{Moullet2023}] lists the following advantages:
\begin{itemize}
    \item Enable new GO science by discovering sources for follow-up investigations, enhancing PRIMA’s pointed science
    \item Make other GO science more efficient by providing flux anchor points for previously undetected sources, allowing optimal exposure time selection, and preventing wasted time on non-detections
    \item Enable novel and unanticipated science for decades into the future
    \item Provide a “deep field” of the Ecliptic Poles, which overlaps with similar fields observed with other space-borne facilities.
\end{itemize}

\begin{figure} [H]
   \includegraphics[width=17.5cm]{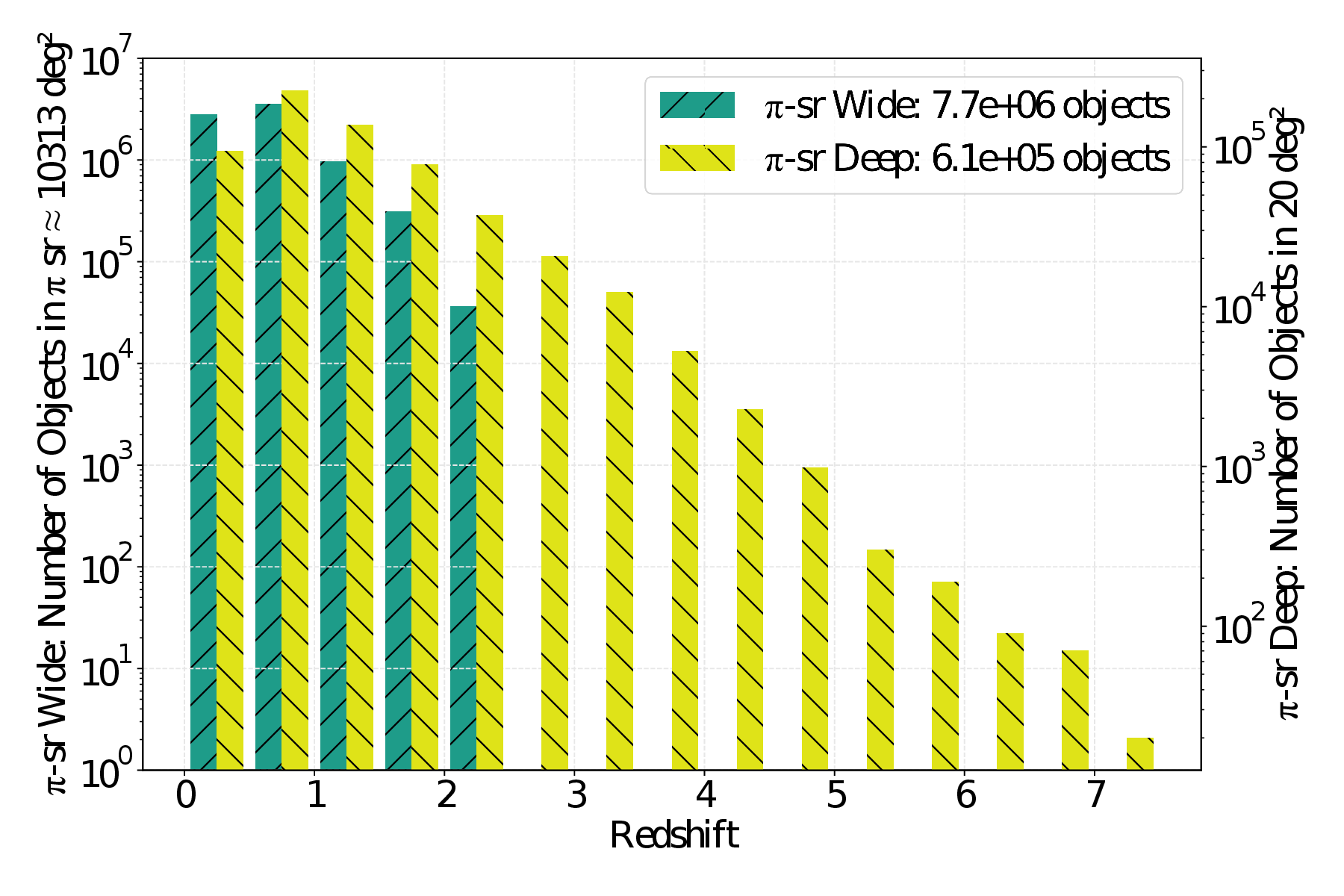}
  \caption{We present the predicted number of objects detected in the $\pi$-IR survey, with the wide part (left and dark green) and the deep part (assumed to be over 20 deg$^2$). We use the Simulated Infrared Dusty Extragalactic Sky (SIDES) semi-empirical simulations and counted all the objects with IR luminosities above the limits (computed using the estimator of Marc Sauvage, private communication) that are given in Tab.~\ref{tab:limits_survey}. Note that the Wide survey axis is to the left and the deep survey axis is to the right. The wide part of the $\pi$-IR survey will detect more than 7 million objects at z $\lesssim$ 2.5 while the deep part will detect objects to z $\gtrsim$ 7.0.}
  \label{fig:pi-IR}
\end{figure}

A wide-area survey will enable significant progress in most if not all the science domains by improving the object statistics, which is the best way to mitigate observational and physical biases. A PHI+PPI survey over 1/4 of the sky, that is $\pi$ sr or, 10313 deg$^2$ could be a well-balanced compromise between statistical coverage and overall exposure time (Figure~\ref{fig:pi-IR}).

\begin{figure} [H]
   \includegraphics[width=18cm]{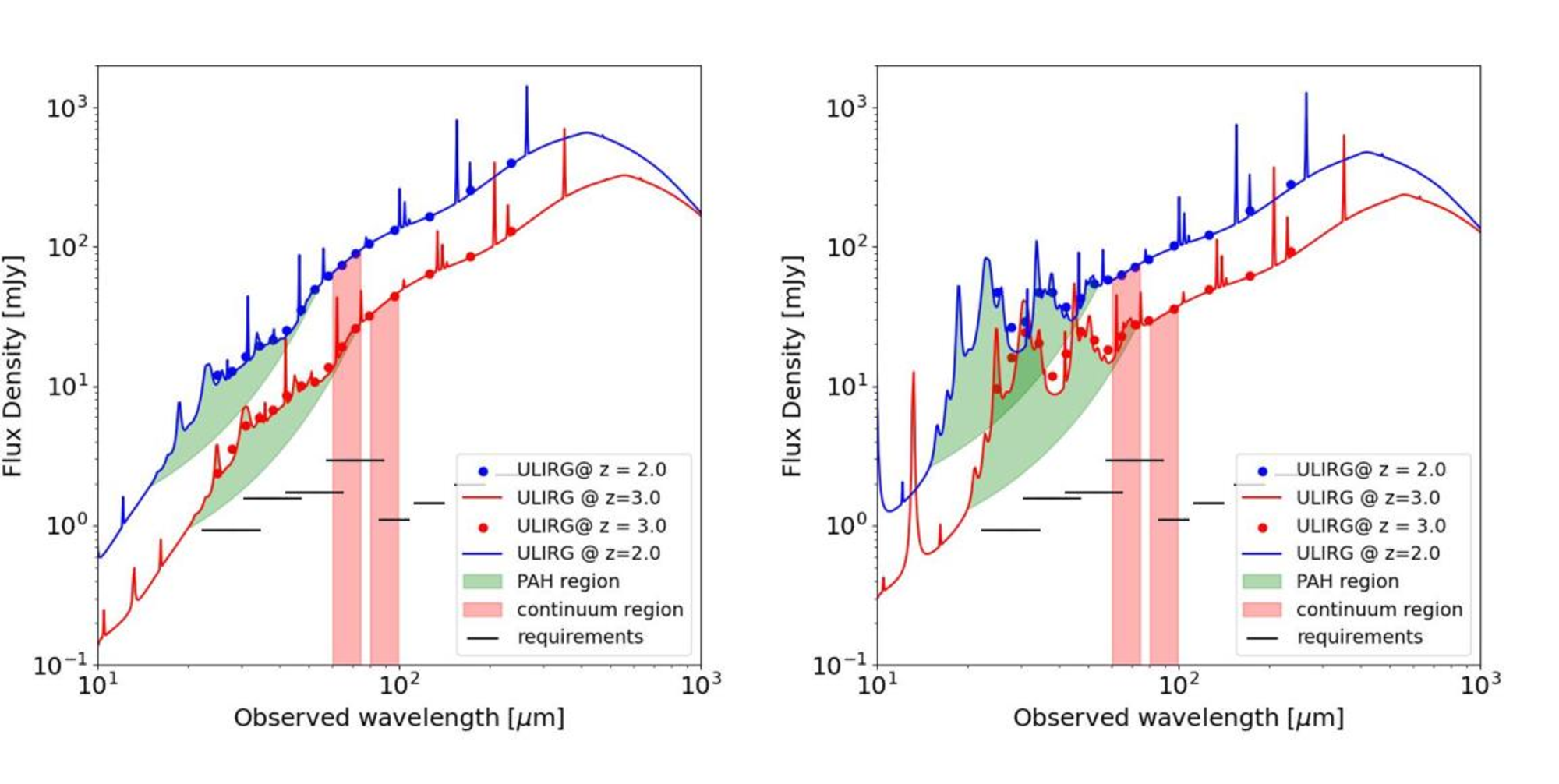}
  \caption{Left: The red curve presents the spectrum of a hyper-LIRG (L$_{\rm dust}$ = 10$^{13}$ L$_\odot$) at z=2 computed by CIGALE, with q$_{PAH}$=0.47. The red dots show where the simulated integrated photometry points will be. However, because PRIMAger's wavelength range starts at 25 $\mu$m, the blue part of the PAH feature will not be observed by PRIMAger. However, the shape of the SED will allow us to measure the redshift. The red curve and the corresponding red dots show that PRIMAger will be able to cover the entire PAH region of a hyper-LIRG at $z=3$. The black horizontal segments show PRIMAger's limiting flux density, that should be reachable with a sophisticated deblending tool that uses priors [\citenum{Donnellan2024}]. Right: Same as left, but with stronger PAH emission, with a PAH mass fraction q$_{PAH}$=7.32.}
  \label{fig:PAH}
\end{figure}

Wright et al. focus on PRIMAger's PPI science. However, PRIMAger's ability to collect data for PHI and PPI simultaneously will allow us to build the first systematic survey of PAH bands by leveraging PRIMAger's hyperspectral mode to derive the strength of the broad PAH features, and complement JWST at higher redshifts. PAH bands form the main tool of many programs [\citenum{Moullet2023}]. PRIMAger PHI mode is able to recover the relative strength of individual PAH features (Fig. \ref{fig:PAH}). Moreover, an important application of PAH features is that they provide us with measurements of the redshift, directly from the observed data. Their strengths are also connected to the gas metallicity as suggested by various works (e.g., Fig.~5 in [\citenum{Schreiber2018}]) and also to the AGN fraction (e.g., Fig.~10 in [\citenum{Bisigello2024}]). 

The $\pi$-IR survey would also be able to cover part of the Galactic plane, in a way similar to Hi-GAL, the Herschel infrared Galactic Plane Survey. Hi-GAL conducted an unbiased photometric survey of the inner Galactic plane between 70 and 500 $\mu$m, aiming at detecting the earliest phases of the formation of molecular clouds and high-mass stars to provide us with a homogeneous census of star-forming regions and ISM. The $\pi$-IR survey would instead focus on the shorter PRIMAger wavelength range, with an overlap between 70 and 235 $\mu$m but with a much better angular resolution in the hyperspectral range and the great advantage of adding the polarimetric information.

With such a large sky coverage, the $\pi$-IR survey will map at comparable sensitivity a wide range of galaxy environments, from voids, filaments, compact and loose groups, up to massive clusters. The survey will thus provide a unique sample of local and high-z galaxies suitable for environmental studies. By comparing objects located in these different density regions, we will be able to quantify the effects of different kinds of interaction, gravitational or hydrodynamical, with the hot gas trapped within the gravitational potential well of groups and clusters and emitting in X-rays, on galaxy evolution [e.g. \citenum{Boselli2006a}; \citenum{Boselli2022}]. The data that PRIMAger will provide are of fundamental importance for measuring the total dust content of galaxies, which can potentially be decreased during any kind of interaction [e.g. \citenum{Longobardi2020a}], and for estimating the total star formation activity of galaxies, generally quenched in high-density regions [e.g. [\citenum{Gavazzi2010}; \citenum{Peng2010}; \citenum{Boselli2014}]. Finally, the exceptional sensitivity of this survey will be optimal for the detection of the diffuse dust component mixed with the intracluster light (ICL) and the intracluster medium (ICM), now observed in a few local massive clusters [e.g. \citenum{Giard2008}; \citenum{gavazzi2016}; \citenum{Longobardi2020b}] and predicted by simulations [\citenum{Gjergo2018}; \citenum{Vogelsberger2019}].

\begin{figure}
\begin{center}
\begin{tabular}{c}
\includegraphics[height=12cm]{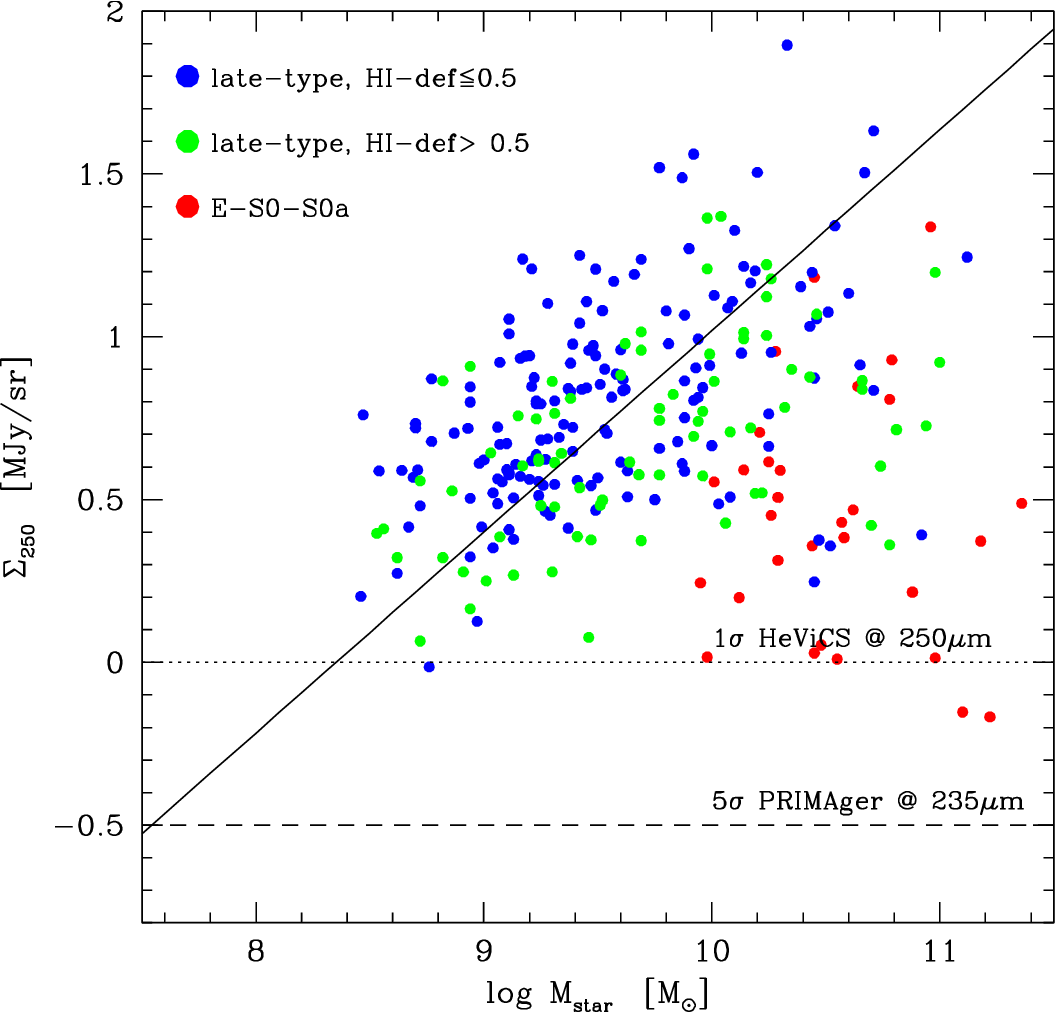}
\end{tabular}
\end{center}
\caption 
{ \label{HRS}
Relationship between 250$\mu$m surface brightness and stellar mass for galaxies in the HRS. Galaxies are coded according to their star formation activity (star-forming vs. quiescent) and atomic gas content (HI-deficient, $HI-{\rm def}$$>$0.5, vs HI-rich, $HI-{\rm def}$$\leq$0.5), here used as a proxy for galaxy environment. The black solid line gives the bisector fit derived for all star-forming systems. The dotted and dashed horizontal lines show the 1$\sigma$ limit of the HeViCS survey in the Virgo cluster and of $\pi$-IR (5$\sigma$), respectively.}
\end{figure}

The results that $\pi$-IR will provide for a sample with unprecedented statistics will be complementary to those obtained for pointed observations of selected objects undergoing different kind of perturbations. The improvement compared to previous surveys conducted with Herschel will be impressive. The Herschel Virgo Cluster Survey (HeViCS, [\citenum{Davies2012}]) mapped the Virgo cluster region over 64 deg$^2$, corresponding to $\simeq$ 1 virial radius, with a 1$\sigma$ sensitivity of $\simeq$ 10, 5, 1 MJy/sr at 100, 160, and 250 $\mu$m, respectively, and detected $\simeq$ 250 galaxies [\citenum{Auld2013}]. We can make a first-order estimate of the number of detected sources in the local Universe. For this purpose, we plot the measured surface brightness at 250 $\mu$m of stellar-mass selected galaxies in the Herschel Reference Survey (HRS), all located at 15$\leq$ $D ({\rm Mpc})$ $\leq$ 25 [\citenum{Boselli2010}] and with flux densities available [\citenum{Ciesla2012}], versus the stellar mass (see Fig. \ref{HRS}). We identify galaxies according to their morphological type and HI-gas content using the HI-deficiency parameter used to distinguish gas-rich systems, typical of the field ($HI-{\rm def}$ $\leq$ 0.5) from gas-poor objects ($HI-{\rm def}$ $>$ 0.5), dominant in clusters [e.g. \citenum{Boselli2006a}]. Those included in the figure are members of the Virgo cluster.

Figure \ref{HRS} clearly shows that star-forming systems of stellar mass $M_{\rm star}$ $\simeq$ 10$^{8.5}$ M$_{\odot}$ can be easily detected with the $\pi$-IR survey in both rich (clusters, groups) and poor (voids, field, filaments) environments. To quantify the possible number of detections we can use the Spring catalog [\citenum{Cattorini2023}] which is composed of all galaxies with a SDSS $r$-band Petrosian magnitude brighter than $r$ $\leq$ 17.7 mag in the sky region 10h $\leq$ R.A. $\leq$ 16h and 0$^o$ $\leq$ Dec $\leq$ 65$^o$ with a recessional velocity $v_{hel}$ $\leq$ 10000 km s$^{-1}$ ($\simeq$ 30000 objects). Considering that the selected sky region is $\sim$ 1/2 that of $\pi$-IR, and that all galaxies in this north Spring catalog have stellar masses $M_{\rm star}$ $\gtrsim$ 10$^{8.2}$ M$_{\odot}$, we can safely predict that the number of detections in the local Universe ($v_{\rm hel}$ $\leq$ 10000 km s$^{-1}$) will be of the order of a few hundred thousands. These objects will span a wide range of environments, from rich clusters such as Coma ($M_{\rm halo}$ $\simeq$ 10$^{15}$ M$_{\odot}$) down to groups, filaments, and voids, thus providing an ideal statistically significant sample of galaxies perfectly suited for environmental studies. At these distances, most of the galaxies have optical sizes $\gtrsim$ 30 arcsec, and are thus resolved by PRIMAger at all frequencies.

Another theme that might benefit strongly from this survey is the identification and analysis of galaxy proto-clusters [\citenum{overzier2016,chiang2017,alberts2022,remus2023}], which Herschel [\citenum{clements2014,clements2016,oteo2018}], Planck [\citenum{planck15,planck16,kubo2019,lammers2022,cheng2019, cheng2020,polletta2021,popescu2023,gatica2024}], the South Pole Telescope [\citenum{rotermund21,hill2022}], JWST [\citenum{laporte2022,Li2024,sillassen2024, morishita2023,morishita2024, polletta2024}], and other facilities have discovered from cosmic noon to cosmic dawn, at redshifts above two, and demonstrated the feasibility of efficiently detecting such structures in the (far-)IR. In the local Universe, about 5 \% galaxies are found in dense clusters of up to thousands of galaxies. Three key questions related to this topic in modern cosmology are how and when (a) such massive structures came together [\citenum{chiang2017,remus2023}], (b) star formation and stellar mass assembly occurred [\citenum{casey2016,polletta2022}], and (c) the baryonic gas efficiently fed these structures [\citenum{daddi2021,daddi2022,daddi2022ApJ}]. Addressing these questions on a statistical basis (unlike the one-at-a-time current approach) will provide insights into the evolutionary processes of galaxy clusters, and the role of dark matter (and interaction with baryons) in the early formation of these cosmic structures.  With its unprecedented combination of size and depth, we anticipate that the {\Large$\pi$}-IR survey will be one of the most impactful projects carried out by PRIMAger; it will yield a powerful legacy database that will be used and cited for more than 20 years, as for IRAS, GALEX, and WISE (and soon Euclid [\citenum{mellier2025,dusserre2025,mai2025}]). and with many more citations than deep-field projects like the Hubble Deep Field (statistics from ADS).

\begin{table} [H]
    \centering
    \begin{tabular}{|l|l|l|l|l|l|l|}
       \hline
Wavelengths [$\mu$m] & 34.3 & 64.5 & 92.0 & 126.0 & 172.0 & 235.0 \\
       \hline
Sensitivity to point sources [mJy] & 2.54e+00 & 3.44e+00 & 7.65e-01 & 1.09e+00 & 1.49e+00 & 2.58e+00 \\
       \hline
Sensitivity to extended sources [MJy/sr] & 1.40e+01 & 5.82e+00 & 6.22e-01 & 4.49e-01 & 3.50e-01 & 3.15e-01 \\
\hline
    \end{tabular}
    \caption{Limiting magnitudes reached with PRIMAger. These values are estimated with the estimator provided by Marc Sauvage (private communication). We assume a maximum scan speed of 250 arcsec/s, and a signal-to-noise ratio SNR=5.0 for the {\Large$\pi$}-IR survey. The number of scan legs is 4657 and the exposure time is 2059 hours -- quite a reasonable investment for such a high-impact project. Wright et al. in [\citenum{Moullet2023}] define an all-sky survey that would take about 5000 hours.}
    \label{tab:limits_survey}
\end{table}

\section{An illustrated picture of PRIMAger's science cases}
\label{sec:sciencecases}
PRIMAger will bring new and unique information to the astrophysical community around the world. We emphasize that the objective of this paper is not to present an exhaustive list of all the projects submitted or feasible with PRIMAger, but to illustrate its capabilities over a wide range of topics  that needs very wide-field observations either because the targetted objects are rare or because large statistical samples are required. We also wish to trigger new projects, ideas and collaborations making use of PRIMA/PRIMAger. 
A first topic is the strong role that could played by PRIMAger on studying highly obscured AGN (Sect.~\ref{sec:AGN}). The build-up of metals and dust (Sect.~\ref{sec:metalsdust}) is the second topic where mid- and far-IR observations with PRIMAger will be very useful to better understand the ISM and its properties. Part of PRIMAger's program will focus on proto-planetary disks, described in the next topic (Sect.~\ref{sec:protoplanetarydisks}).  In the much more local Universe, PRIMAger will also collect important information on small bodies in the Solar System (Sect.~\ref{sec:kuiperbelt}). 

\subsection{Complementarity of X-ray and AGN surveys for Compton-thick AGN}
\label{sec:AGN}

Recent work suggests that the fraction of obscured AGN could be as high as about 80\% at z $\sim$ 6 [\citenum{Vito2018}]. This might be due to the compactness and high density of gas within the host galaxies. These obscured AGN are also needed for X-ray background synthesis models: a large fraction of the yet unresolved X-ray background could be due to the most obscured Compton thick AGN, with N$_H$ $>$ 10$^{24}$ cm$^{-2}$.

An AGN study would benefit from the synergy between PRIMA and ATHENA: it needs mid- and far-IR observations with enough spectral resolution in imaging to extract AGN diagnostics for the PAH emission at z$>$1 [e.g., \citenum{Li2024}] in complement to X-ray observations. 

The main information necessary to derive quantitative diagnostics on AGN is the possibility to fit and separate the prominent PAH features. Contraints in the PAH emission bands and the underlying dust continuum with, e.g. a fit with the CIGALE code, mean that we can separate and measure the different bands. PRIMAger R$\sim$8 spectral resolution (Tab.~\ref{tab:prima_instruments}) is capable of performing this task, especially at z $>$ 1.5 when the all brightest PAH bands are observed with PRIMAger.

\begin{figure} [H]
   \begin{center}
   \begin{tabular}{c} 
   \includegraphics[width=17cm]{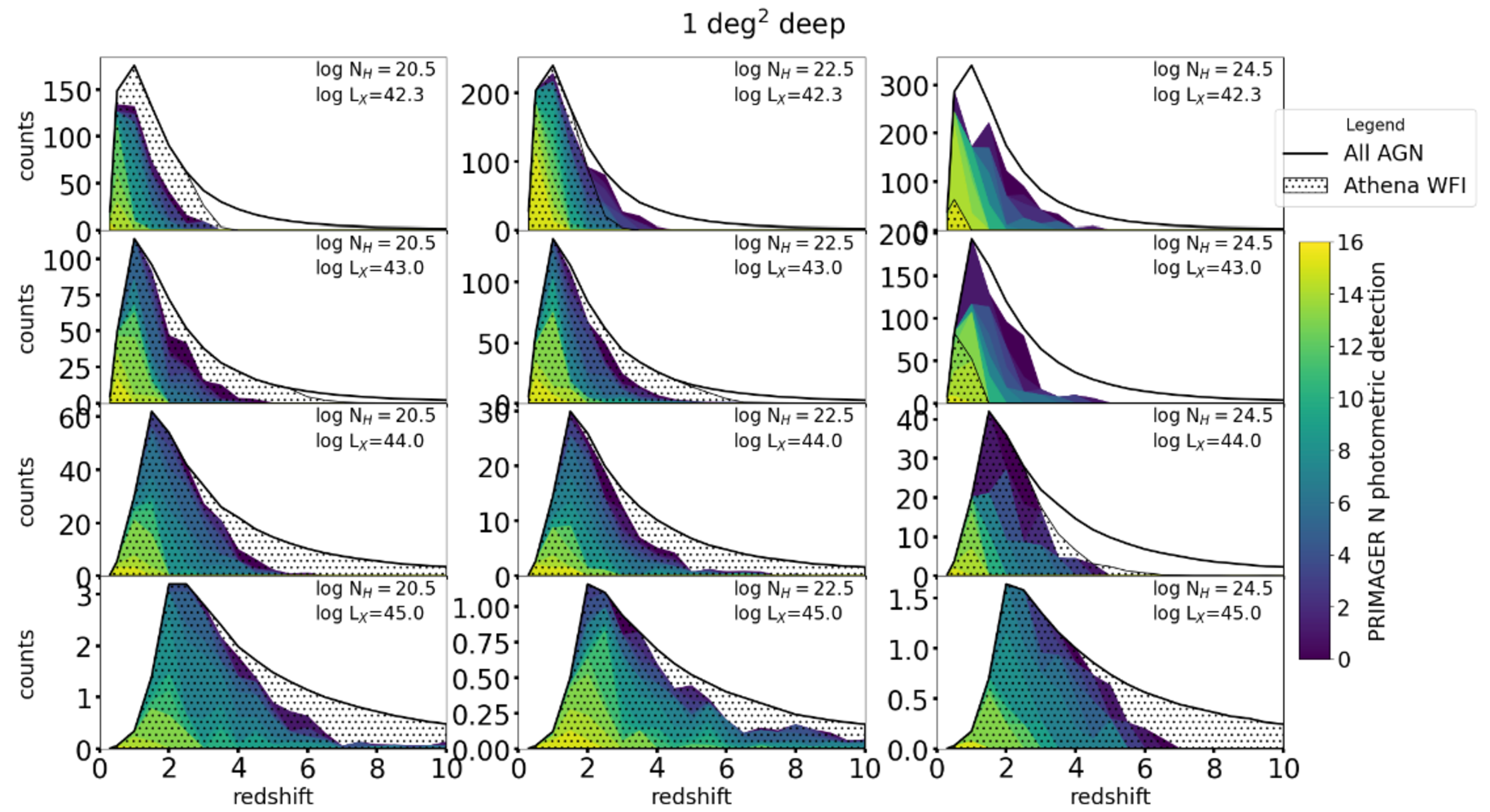}
   \end{tabular}
   \end{center}
   \caption[example] 
   {\label{fig:AGN} Figure from the Compton-thick AGN science case developed by Barchiesi et al. in [\citenum{Moullet2023}]: Based on their simulations, PRIMAger will detect about 6000 Compton-thick AGN at z$>$6. Low- and mid-luminosity Compton-thick AGN are not detected in the current and next generation of X-ray and ultraviolet surveys (top-right of the plot). PRIMAger will complement X-ray surveys by observing this population of heavily obscured AGN, and bridge the gap between the predicted and the observed black hole accretion rate.}
\end{figure}

Most scenarios assume that high-accretion rates onto black holes occur in highly dust-obscured conditions, implying that the ultra-violet through optical emission is likely to be extremely difficult to observe. Coupling IR and X-rays uniquely exploits both the X-ray and IR bands to detect AGN over the whole parameter space: ATHENA will detect high-luminosity AGNs (L$_X$$>$10$^{44}$ erg~s$^{-1}$), even at high redshifts, while PRIMAger will uncover highly-obscured ones at redshifts z$>$2 (Fig.~\ref{fig:AGN}).

Conversely, if only one of the two instruments—either PRIMA or ATHENA—is available, any AGN study would suffer from selection bias due to incomplete sample coverage. PRIMA, in particular, plays a crucial role: without its sensitivity in the mid- and far-infrared, low- and mid-luminosity Compton-thick AGN would go undetected, leading to a biased census. Moreover, as shown in Fig.~\ref{fig:AGN}, the absence of infrared observations would introduce a significant bias against dust-enshrouded phases of the evolution of AGN. This would prevent a comprehensive understanding of AGN growth and obscuration throughout cosmic time. If the $\pi$-IR survey is actually performed, it could be defined to contain this program.

\subsubsection{Co-evolution of Dust and Metals}
\label{sec:metalsdust}

PRIMAger will provide us with a very useful probe (Fig.~5 in [\citenum{Schreiber2018}]) of metallicity (Z$_{\rm ISM}$) and dust mass (M$_{\rm dust}$) from PAH bands and fine-structure line measurements (MeDALiC program by Burgarella et al. in [\citenum{Moullet2023}]). Previous IR observatories (AKARI, Spitzer, Herschel in space, and ALMA, NOEMA on the ground) lack the sensitivity to carry out the former measurements  and lack statistics for the latter. PRIMAger will bring both, especially if a very wide sky PRIMAger survey is carried out. Such a survey will combine a wide 10,000 sq. degrees survey with a deep survey in the ecliptic poles, which will be used to probe the evolution of PAH emission traced by PRIMAger's hyperspectral channel at higher redshifts. 

\begin{figure} [H]
\begin{centering}
     \includegraphics[width=12.0cm]{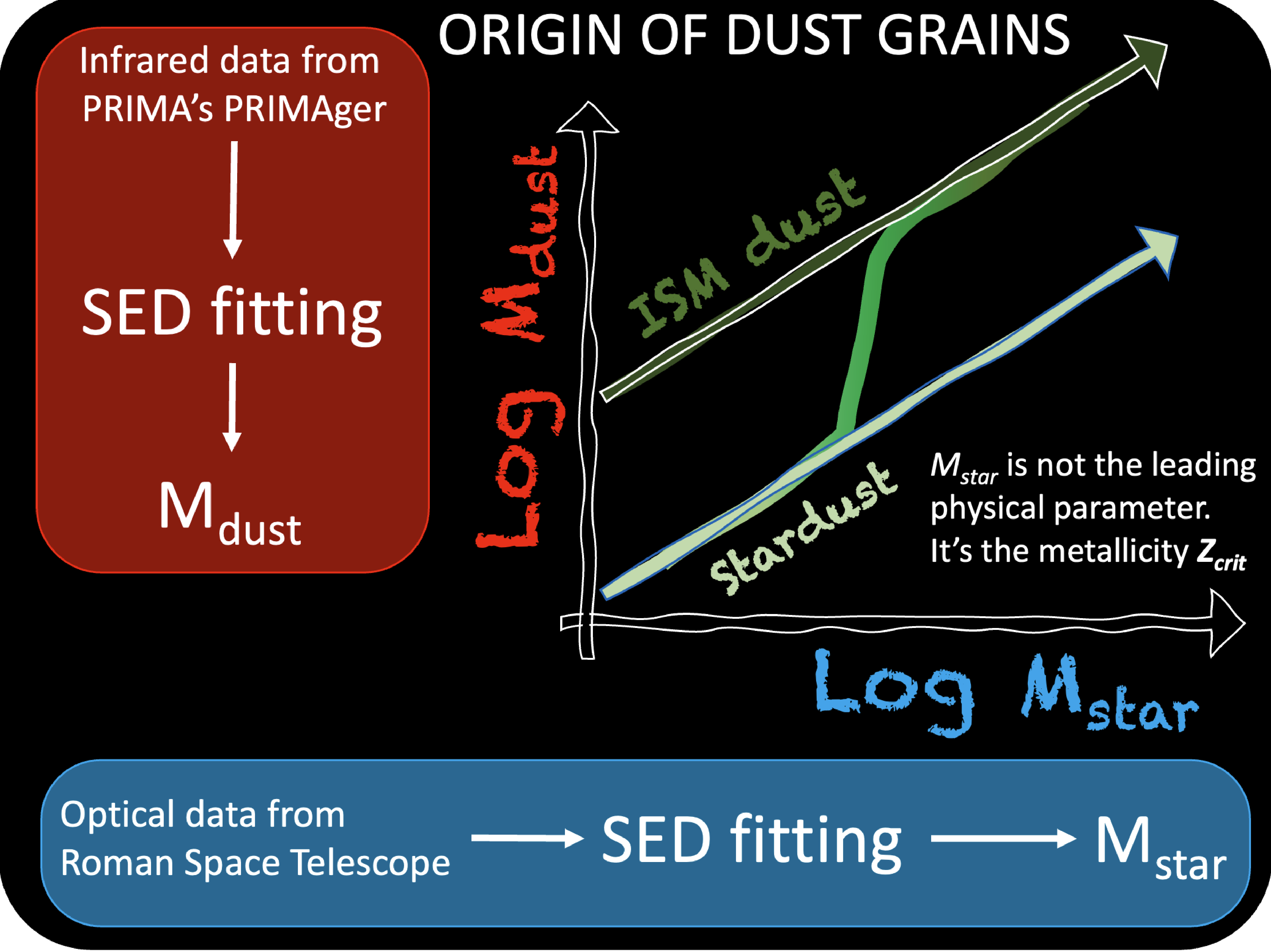}
  \caption{M$_{dust}$ as a function of M$_{star}$. PRIMAger will provide us with the mid and far-IR data for a large sample of galaxies from which we will estimate the dust masses via SED fitting. For a large part of the PRIMAger sample, Roman will secure optical and near-IR data from which we will derive stellar masses via SED fitting. Positioning the galaxies in the M$_{dust}$ versus M$_{star}$ diagram will allow us to derive the main origin of the dust mass in these galaxies: if a galaxy lies on the upper dark-green line [\citenum{Witstok2023}], the dust mass is dominated by ISM-grown grains. If the galaxy lies on the bottom light-green line, the dust mass of the galaxy is dominated by stardust [\citenum{Witstok2023}] that underwent a 95 \% destruction of grains by reverse SNe shock. In between these two lines, we will find transitional objects. Massive galaxies are expected to be on the top sequence while lower-mass galaxies ($\log_{10}$ (M$_{star}$) $\lesssim$ 9.0 are mainly located in the bottom sequence [\citenum{Burgarella2025}]. However, models suggest that the leading parameter that triggers the transition is not directly the stellar mass but it is the metallicity [\citenum{Inoue2011, Graziani2020, Triani2020, Parente2022, Choban2024}]}
  \label{fig:MdustMstar}
\end{centering}
\end{figure}

Dust plays a critical role in the formation of low-mass stars by facilitating several key processes in the interstellar medium (ISM): collisions between gas and dust grains enable efficient gas cooling, particularly at high densities (n$_H$ $\sim$ 10$^{12}$ cm$^{-3}$), which promotes fragmentation of the cloud and the formation of low-mass stars. Later generations of stars added their own production of metals and dust to the ISM of galaxies via asymptotic giant branch (AGB) stars. These dust grains are also fundamental to form planetary systems, through complex physical processes. After this first phase that seeds the ISM, the growth of the dust mass of most galaxies is thought to be produced by accretion of metals on grains seeds in the ISM, enhancing the dust mass [\citenum{Zhukovska2014, Asano2013}]. This process only happens when a critical ISM metallicity is reached at 0.05 $<$ Z/Z$_\odot$ $<$ 0.5 [\citenum{Inoue2011, Graziani2020, Triani2020, Parente2022, Choban2024}]. A first mandatory phase is thus the production of seed dust grains (stardust) in the circumstellar medium of supernovae, and later AGB stars, that is followed by a transition to the more efficient dust mass growth. This transition is thought to be observed both in the very local universe [\citenum{Remy-Ruyer2014}] and at z$>$4 to the earliest cosmic time [\citenum{Burgarella2025}]. 
PRIMA will address fundamental question of the life cycle of dust by measuring the dust mass from PRIMAger's spectral energy distributions via fitting and stellar mass by using rest-frame near-IR data from Roman and Euclid. The location of galaxies in the M$_{dust}$ versus M$_{star}$ diagram (Fig.~\ref{fig:MdustMstar}) will provide us with a tracer of the origin of the dust mass. Two different sequences are predicted by models [\citenum{Graziani2020 , Mancini2015, Esmerian2024, Witstok2023}] and now observed [\citenum{Burgarella2025}].

The detection of such stardust galaxies that become rare at z $\lesssim$ 6 - 7 requires the observation of large galaxy samples to monitor the evolution of this dust cycle with cosmic time. A {\Large $\pi$}-IR survey is thus the best method required to identify such objects, and to study them.

\begin{figure}
    \centering
   \includegraphics[width=7.0cm, angle=-90]{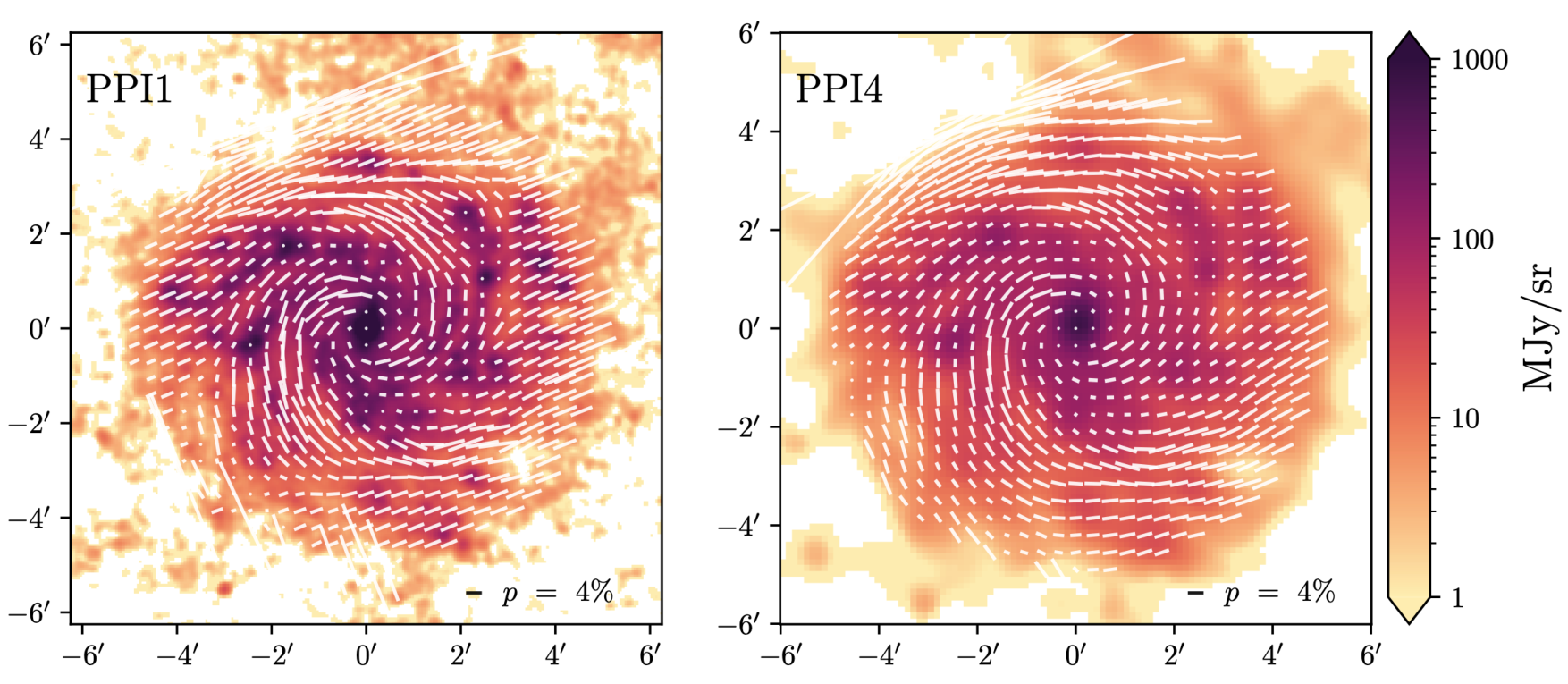}
  \caption{Figure extracted from [\citenum{Dowell2024}]: the polarization vectors correspond to the magnetic field orientation in a simulation for the galaxy NGC 6946. They are overlayed on the total intensity maps for PPI1 ($\lambda_{\rm central}$ = 96.3 $\mu$m), left) and PPI4 ($\lambda_{\rm central}$ = 235 $\mu$m, right). The vector length is proportional to the polarization fraction.}
  \label{fig:polarization}
\end{figure}

B\'ethermin et al. [\citenum{Moullet2023}] present a program that combines two different 200h: a deep (2 deg$^2$) and a medium-wide (20 deg$^2$) field. These surveyed areas have been selected to provide about 10,000 detections up to z = 2 while no detections have been published so far beyond the nearby Universe. SIDES [\citenum{Bethermin2022}] semi-empirical simulation modified to include polarization, was used. This number will allow us to test which properties (inclination, stellar mass, gas density, turbulence, outflows, presence of a starburst and/or AGN, environment) drive the dust polarization. The depth will not be affected by the confusion limit in polarization. PRIMAger's long-wavelength polarimetric bands will study how the structure of interstellar dust grains changes across environments in galaxies. A detailed analysis [\citenum{Dowell2024}] describes the approach that PRIMA can take to build ultra-deep maps of intensity and polarization in four bands in the 91-232 $\mu$m wavelength range. This approach permits measurements of Stokes I, Q, and U in single scans. Dust polarization holds critical clues about the origin of dust grains and how they are processed in the ISM of galaxies (Fig.~\ref{fig:polarization}). Dying carbon-rich asymptotic giant branch stars produce unpolarized dust that is ejected into the ISM. If the ISM is composed of ejected stardust, we expect low polarization. However, theoretical models suggest an efficient growth of dust in ISM and composite grains grown by aggregating stardust with ISM-grown dust do produce polarization [\citenum{HensleyDraine2023}]. This process is expected to become efficient when the metallicity reaches a critical value, which is the metallicity at which the stardust mass is equivalent to the growth of the dust mass in the ISM [\citenum{Asano2013}]. If the $\pi$-IR survey is actually performed, it could be defined to contain this program.


\begin{figure} [H]
   \begin{center}
   \begin{tabular}{c} 
   \includegraphics[width=12cm]{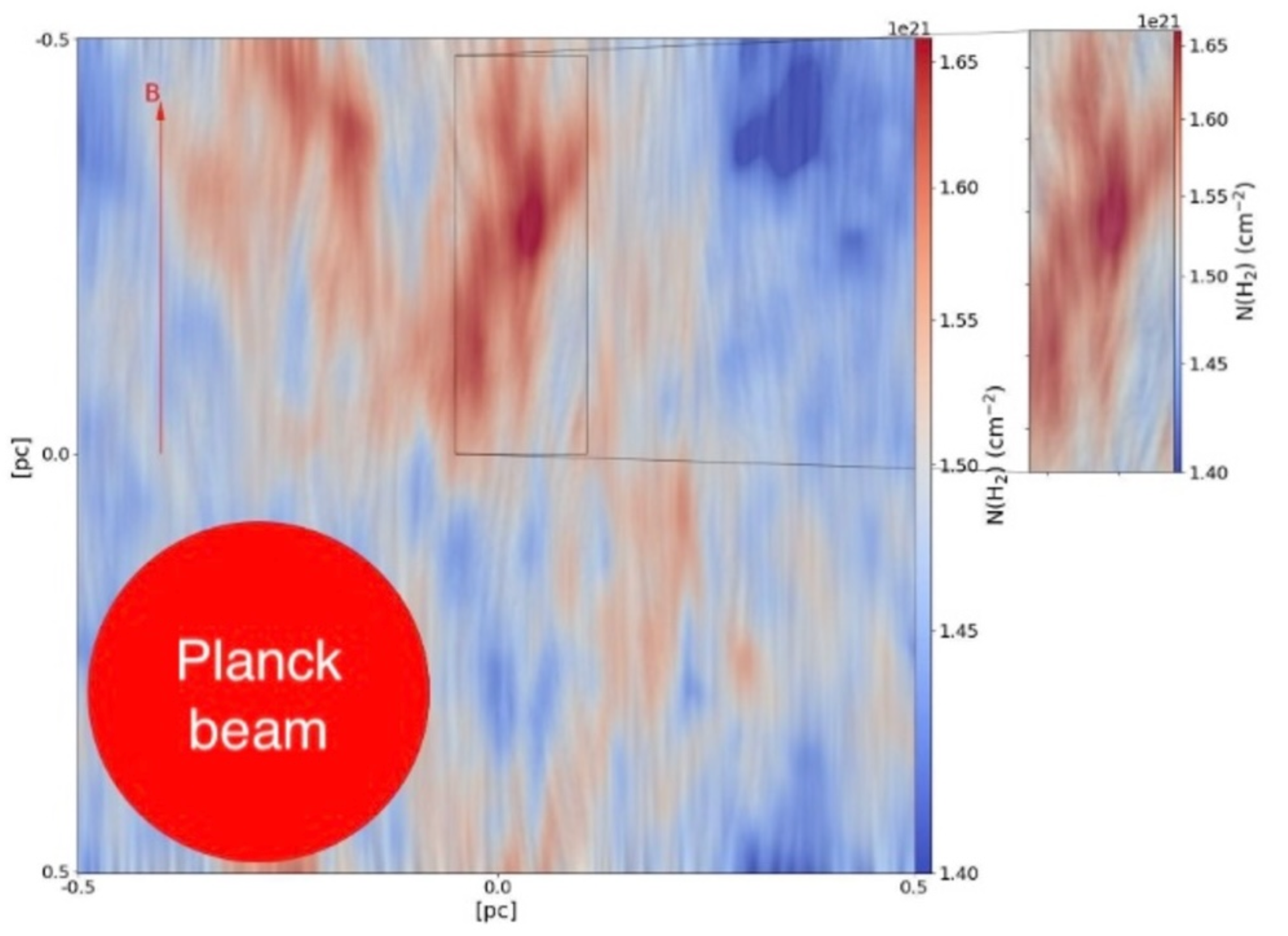}
   \end{tabular}
   \end{center}
   \caption[Figure extracted from Tassis et al. in [\citenum{Moullet2023}]: Striations in a 3D MHD simulation of a cloud. The background colormap shows the column density of the simulated cloud and the structured pattern shows the magnetic field lines, both convolved to PRIMAger’s resolution at 235 $\mu$m. MHD wave effects should be observable in these regions as different wave modes imprinted in the striations pattern. The zoomed-in panel shows the passage of a small-scale MHD mode that propagates along the mean field orientation. Right: The zoomed-in panel shows the passage of a "sausage mode”. We stress that it is not possible to carry this type of work out so far. Previous far-IR facilities such as Planck could not reach the 0.1 pc resolution (15 arcsec at the distance of Taurus) necessary to probe such striations.] 
  \label{fig:striations}
\end{figure}

Tassis et al. [\citenum{Moullet2023}] show that the information brought by PRIMAger on the polarization, and thus on the magnetized structures in the outskirts of ISM clouds, is a strong asset for understanding the elongated hair-like and diffuse structures observed in interstellar molecular clouds. These linear striations encode information on early phases of star formation processes (Fig.~\ref{fig:striations}). This program tests whether these striations are formed by magnetohydrodynamic (MHD) wave-modes. Only PRIMAger provides us with both the polarimetric capabilities and the spatial resolution to reach 0.1-pc scales, inaccessible so far with other IR facilities. Combining this information with the stellar polarization information collected in the optical will open the door to a brand new way to explore the properties of dust grains in the ISM.

\subsection{Solar System's Kuiper-Belt Objects}
\label{sec:kuiperbelt}

PRIMAger will also be useful for studying the Solar System (Moullet et al. in [\citenum{Moullet2023}]) via the size frequency distribution of Kuiper-Belt objects (KBOs). Because the Kuiper Belt is thought to be the most easily observable example of the last stage of an evolved debris disk, the composition and size of individual Kuiper Belt Objects can be linked to the original composition
of the Outer Solar System. This will provide information about the formation history of the outer Solar System through a comparison to theoretical models, as it covers the 40–100 km region where models display a slope break in the distribution; observing the regime below 100 km constrains the role of destructive and accretive processes that modified the initial planetesimal mass function. From PRIMAger observations at 100 $\mu$m, it will be possible to unravel the effect of  albedo in retrieving the equivalent size and to detect objects in the 35–80 km diameter range (Fig.~\ref{fig:KBOs}), inaccessible to Herschel and Spitzer.

{
\sidecaptionvpos{figure}{t}
\begin{SCfigure}
   \includegraphics[height=8cm]{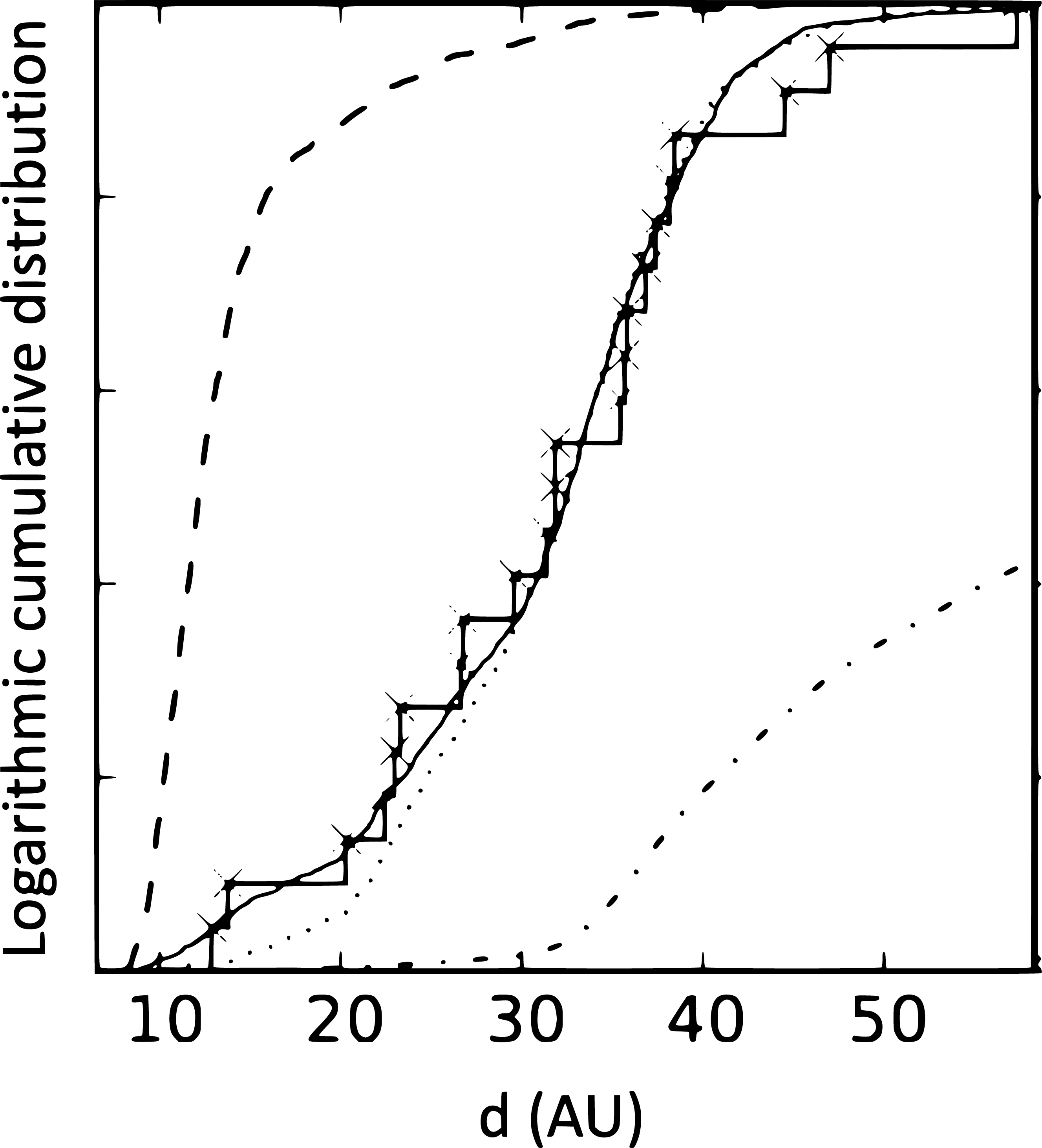}
  \caption{Extracted from Shankman et al. (2016). Using the good far-IR point-source sensitivity offered by PRIMAger at 100 $\mu$m (about 100 times better than Herschel-PACS), a pointed survey of 100–200 KBOs will focus on the size regime of objects down to 35 km diameter, which is of interest for theoretical models. The survey will provide us with the photometric data that will be converted to equivalent sizes, using available optical magnitudes and thermal emission models. The cumulative distribution of KBOs scattered population as a function of size (red - based on Outer Solar System Origins Survey observations) are compared to three modeled distributions. The albedo assumption is 5\% for all bodies.}
  \label{fig:KBOs}
\end{SCfigure}
}

\subsection{Protoplanetary Disks}
\label{sec:protoplanetarydisks}

\begin{figure} [H]
   \begin{center}
   \begin{tabular}{c} 
   \includegraphics[width=10cm]{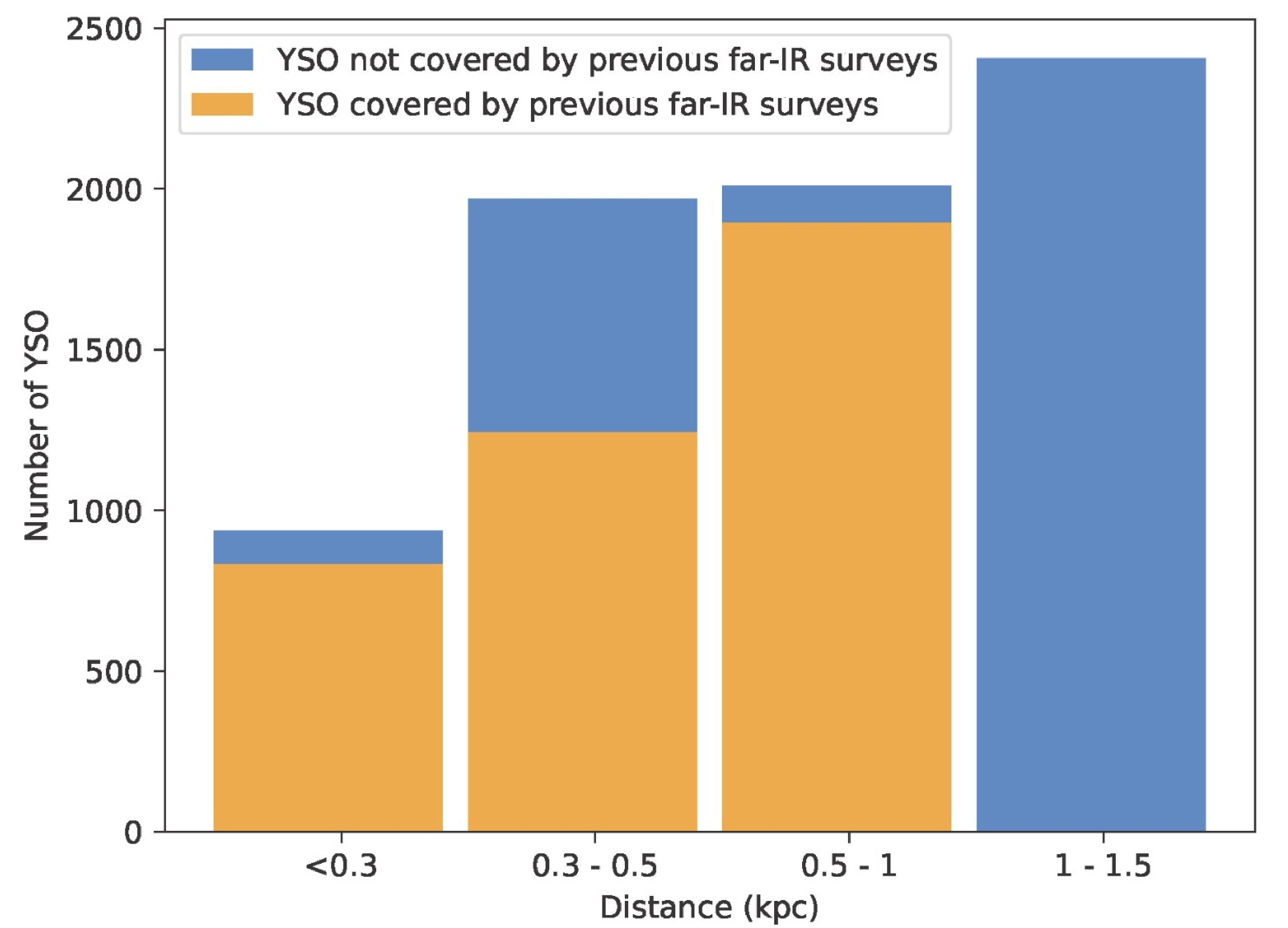}
   \end{tabular}
   \end{center}
   \caption[example] 
   {\label{fig:protoplanetary} Extracted from Villenave et al. in \citenum{Moullet2023}. Estimate of the number of young stellar objects (YSO) of all evolutionary stages included in regions previously covered or not covered by far-IR surveys. PRIMA will double the number of YSO with far-IR fluxes and expand statistics on disk properties over a wider range of stellar ages and external environments.}
\end{figure}

PRIMAger will shed new light on how protoplanetary disks evolve (Villenave et al. in [\citenum{Moullet2023}]). Planet formation is a complex process that is but one part of the evolution of protoplanetary disks. To characterize and analyze this evolution and its properties, a statistical approach is fundamental. However, the small number of existing far-IR observations hampers the derivation of significant statistical trends. The mid-IR emission allows the observations of the innermost regions of disks, while the far-IR is most sensitive to the vertical structure of the disk to characterize the grain size and mass of dust, a crucial element in forming disks. This proposed program aims at doubling the number of far-IR observed disks, and covers a wider range of star-forming environments. Increasing the size of the sample by a factor of two will open up the analysis to a wider range of star-forming environments by observing older regions and trace the evolution of disks, about also higher-mass star-forming regions to trace different UV environments. Ultimately, this would allow us to study their dependence to parameters such as the stellar spectral type, age, mass, the binarity of the system, or the disk location within a high mass or low mass star-forming region that were unacessible due to the low statistics. The high sensitivity of PRIMAger could collect far-IR photometric observations of all star-forming regions up to 1.5 kpc in which 9 regions of interest, covering sky areas between 1 deg$^2$ and 100 deg$^2$ have been identified. The SEDs for thousands of disks (Fig.~\ref{fig:protoplanetary}) will thus be completed. Note that previous far-IR surveys did observe protoplanetary disks and young stellar objects with observations between 70 and 500 $\mu$m of large cloud areas up to 3 kpc. However, they were targeting very young stellar objects not cover older star-forming regions within 1.5 kpc. By widening the range of star-forming environments and observing older regions, we can trace the evolution of disks and higher-mass star-forming regions in various UV environments.

\section{PRIMAger's confusion}
\label{sec:confusion}

\begin{figure} [H]
   \begin{center}
   \begin{tabular}{c} 
   \includegraphics[width=14cm]{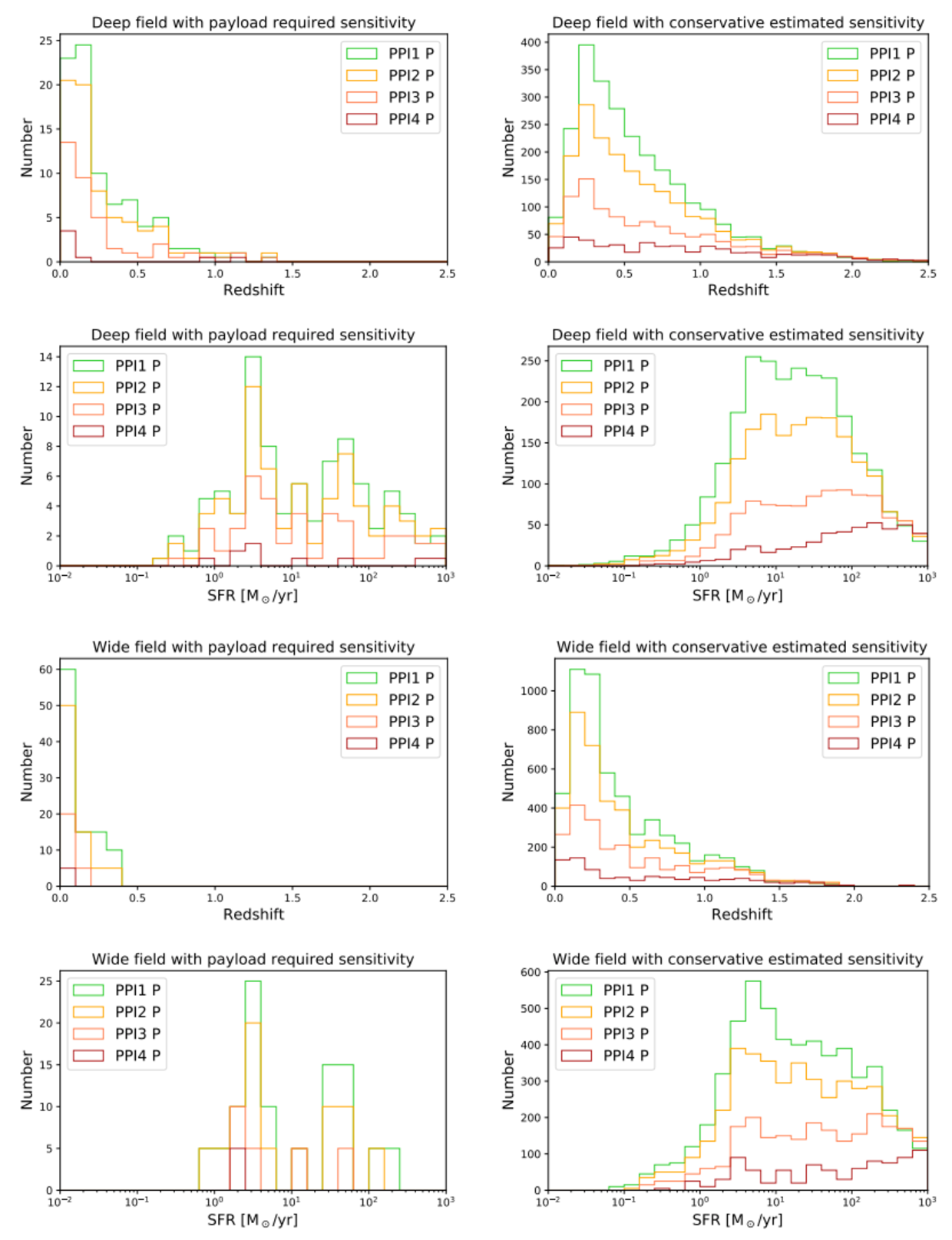}
   \end{tabular}
   \end{center}
   \caption[example] 
   {\label{fig:Bethermin24}Figure adapted from [\citenum{Bethermin2024}]: redshift and SFR distributions of the sources above the detection limit in polarized flux density. The left columns correspond to the payload required sensitivity, and the right ones to the conservative estimated sensitivity. From top to bottom, the rows are: redshift distribution in the deep field (5$\sigma$ polarized flux density of about 0.3 mJy in PPI-1), SFR distribution in the deep field, redshift distribution in the wide field, (5$\sigma$ polarized flux density of about 1.0 mJy in PP1-1) SFR distribution in the wide field.}
\end{figure}

\begin{figure} [H]
   \begin{center}
   \begin{tabular}{c} 
   \includegraphics[angle=-90,width=17cm]{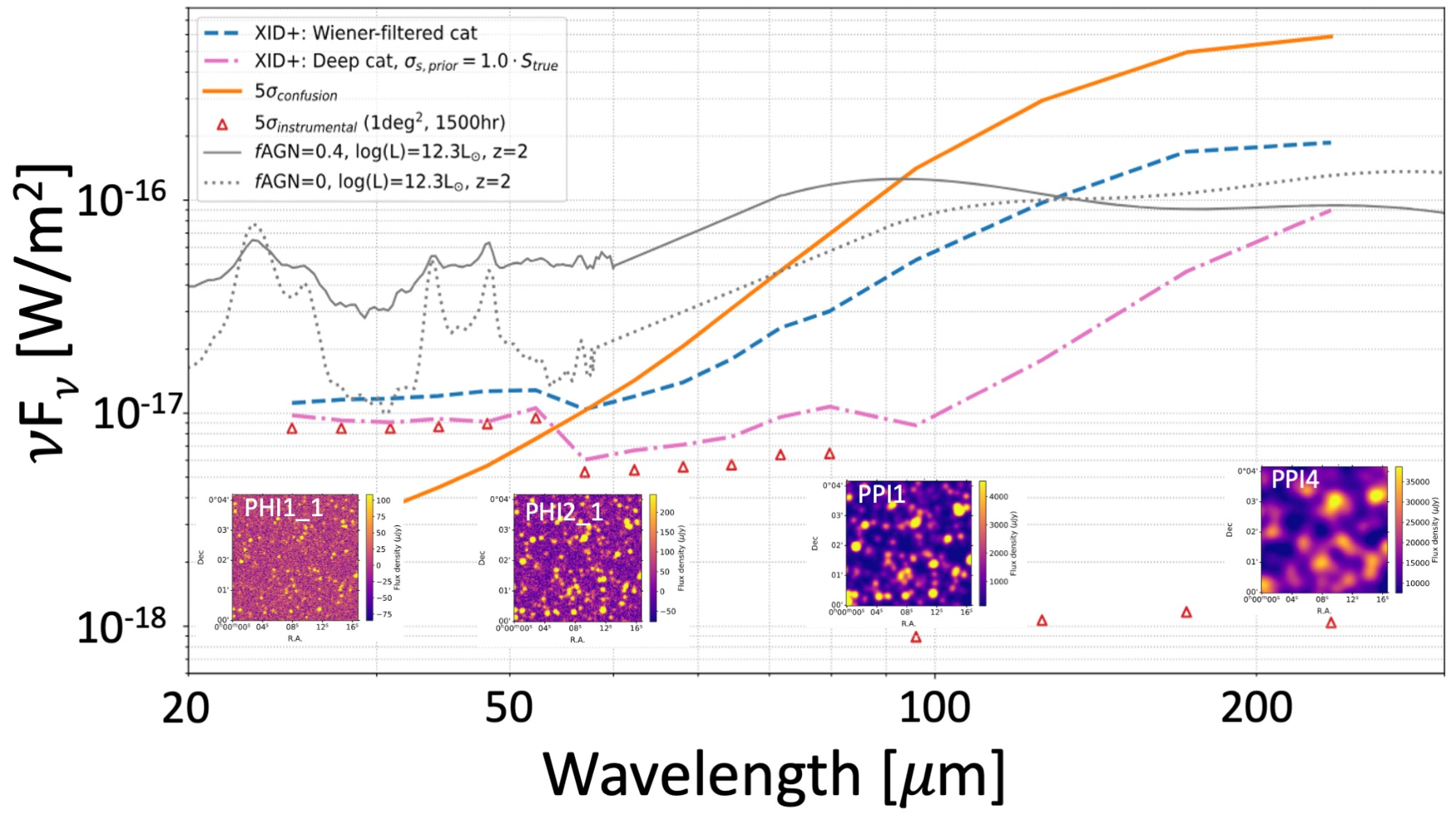}
   \end{tabular}
   \end{center}
   \caption[example] 
   {\label{fig:Donnellan23}Figure adapted from [\citenum{Donnellan2024}]: the limiting flux density as a function of wavelength from 25–235 $\mu$m for XID+ deblending. Using positional and weak flux priors allow us to reach a 5$\sigma$ depth, that is more than an order of magnitude fainter than the classical confusion limit at wavelengths $>$ 100 $\mu$m. Moreover, using XID+ allows the measurement of SEDs for a typical galaxy at $z = 2$ to $\lambda$ = 126 $\mu$m with only positional priors and out to the longest PRIMAger PPI channel with the addition of a weak intensity prior.}
\end{figure}

When working on a space-based mid- and far-IR project, the combination of wavelength, costs and launch constraints translate into limited angular resolution. In other words, in order to extract the science from these data, one must overcome the problem of confusion. Confusion is the blending of several objects (most often galaxies) that lie within the same instrumental beam (sometimes in physical proximity, and sometimes physically unassociated, but along the line of sight). Consequently, only bright objects can be identified, and they may include some contribution from nearby fainter and not individually detected sources. Moreover, confusion increases rapidly with wavelength [\citenum{Bethermin2024}]. 
The PRIMAger team has quantified how this confusion impacts the performance of basic blind source extractors, both in intensity and polarization and finally in the measurement of physical parameters [\citenum{Bethermin2024}; \citenum{Donnellan2024}]. The observations simulated in these papers are produced with the classical confusion limit for all PRIMA bands (Fig.~\ref{fig:Bethermin24}) using the simulated infrared extragalactic sky semi-empirical simulation (SIDES, \citenum{Bethermin2022}). The PRIMAger team has also demonstrated that the flux modelling capabilities of the code XID+ [\citenum{Hurley2017}] provide flux measurements of galaxies in the PRIMAger simulated maps that reach below the classical confusion limit (Fig.~\ref{fig:Donnellan23}). 

\section{Conclusions}
\label{sec:conclusion}

PRIMA and PRIMAger will provide us with an exceptional observatory, hopefully by the beginning of the next decade, that will benefit the entire astrophysical community, from the Solar System to the distant Universe, at redshifts close to the Epoch of Reionization. Combined with the multi-wavelength data collected by other wide-area surveys (GALEX, Euclid, the Roman Space Telescope) and with the extension in X-rays brought by New-Athena from the European Space Agency, the combination of the wavelength range, the mapping speed and the relatively high spatial resolution will make PRIMAger on PRIMA a truly transformational instrument. 

A {\Large$\pi$}-IR survey would take about 2000 hours of PRIMA's total observing time of 5 years to observe a quarter of the infrared sky from 26 to 260 $\mu$m with PRIMAger's unique hyperspectral and polarimetric modes. Combined with PRIMAger's FIRESS spectroscopic instrument, {\Large$\pi$}-IR will provide an unprecedented rich legacy dataset for both Galactic and extragalactic survey science.

Only one very-wide or all-sky survey is likely to be carried out by PRIMA. This means that a wide and international collaboration would certainly be the best approach to optimize the preparation, observation, data processing and release of such a fundamental effort.

\section*{ACKNOWLEDGMENTS}       
 
A portion of this research was carried out at the Jet Propulsion Laboratory, California Institute of Technology, under a contract with the National Aeronautics and Space Administration (80NM0018D0004). We also acknowledge support from the French Centre National d'Etudes Spatiales (CNRS).
This paper presents an extension and an update of a SPIE paper [\citenum{Burgarella2024_SPIE}]. The authors thank Rachel Somerville for reading and commenting a preliminary version of this paper.

\section*{DISCLOSURES}

The authors declare there are no financial interests, commercial affiliations, or other potential conflicts of interest that have influenced the objectivity of this research or the writing of this paper

\section*{CODE AND DATA AVAILABILITY}
No specific data were used for this paper. To simulate the observations in Sect.~\ref{sec:allsky}, we used CIGALE which is an open-access code available from http://cigale.lam.fr

\bibliography{Article} 
\bibliographystyle{spiebib} 

\end{document}